\documentclass[prl,twocolumn,
superscriptaddress,
 amsmath,amssymb,
 aps,floatfix
]{revtex4-2}

\usepackage{gensymb}
\usepackage{graphicx}
\usepackage{dcolumn}
\usepackage{bm}
\usepackage{multirow}
\usepackage{textcomp, gensymb}
\usepackage{xcolor}
\usepackage
{}

\newcommand{\fgt}{Fe$_3$GeTe$_2$}
\newcommand{\fgat}{Fe$_3$GaTe$_2$}

\newcommand{\tc}{T$_C$}
\newcommand{\ef}{$E_F$}

\begin{document}

\title{Spectral Evidence for Local-Moment Ferromagnetism in van der Waals Metals Fe$_3$GaTe$_2$ and Fe$_3$GeTe$_2$}

\author{Han Wu}
\affiliation{Department of Physics and Astronomy and Rice Center for Quantum Materials, Rice University, Houston, TX, 77005 USA}

\author{Chaowei Hu}
\affiliation{Department of Physics, University of Washington, Seattle, Washington 98195, USA}
\affiliation{Department of Materials Science and Engineering, University of Washington, Seattle,
Washington 98195, USA}

\author{Yaofeng Xie}
\affiliation{Department of Physics and Astronomy and Rice Center for Quantum Materials, Rice University, Houston, TX, 77005 USA}

\author{Bo Gyu Jang}
\affiliation{Theoretical Division and Center for Integrated Nanotechnologies, Los Alamos National Laboratory, Los Alamos, NM, USA}

\author{Jianwei Huang}
\affiliation{Department of Physics and Astronomy and Rice Center for Quantum Materials, Rice University, Houston, TX, 77005 USA}

\author{Yucheng Guo}
\affiliation{Department of Physics and Astronomy and Rice Center for Quantum Materials, Rice University, Houston, TX, 77005 USA}

\author{Shan Wu}
\affiliation{Department of Physics, University of California at Berkeley, Berkeley, California 94720, USA}

\author{Cheng Hu}
\affiliation{Advanced Light Source, Lawrence Berkeley National Laboratory, Berkeley, CA 94720, USA}

\author{Ziqin Yue}
\affiliation{Department of Physics and Astronomy and Rice Center for Quantum Materials, Rice University, Houston, TX, 77005 USA}

\author{Yue Shi}
\affiliation{Department of Physics, University of Washington, Seattle, Washington 98195, USA}

\author{Rourav Basak}
\affiliation{Department of Physics, University of California San Diego,La Jolla, CA 92093,USA}

\author{Zheng Ren}
\affiliation{Department of Physics and Astronomy and Rice Center for Quantum Materials, Rice University, Houston, TX, 77005 USA}

\author{T. Yilmaz}
\affiliation{National Synchrotron Light Source II, Brookhaven National Lab, Upton, New York 11973, USA}

\author{Elio Vescovo}
\affiliation{National Synchrotron Light Source II, Brookhaven National Lab, Upton, New York 11973, USA}

\author{Chris Jozwiak}
\affiliation{Advanced Light Source, Lawrence Berkeley National Laboratory, Berkeley, CA 94720, USA}

\author{Aaron Bostwick}
\affiliation{Advanced Light Source, Lawrence Berkeley National Laboratory, Berkeley, CA 94720, USA}

\author{Eli Rotenberg}
\affiliation{Advanced Light Source, Lawrence Berkeley National Laboratory, Berkeley, CA 94720, USA}

\author{Alexei Fedorov}
\affiliation{Advanced Light Source, Lawrence Berkeley National Laboratory, Berkeley, CA 94720, USA}

\author{Jonathan Denlinger}
\affiliation{Advanced Light Source, Lawrence Berkeley National Laboratory, Berkeley, CA 94720, USA}

\author{Christoph Klewe}
\affiliation{Advanced Light Source, Lawrence Berkeley National Laboratory, Berkeley, CA 94720, USA}

\author{Padraic Shafer}
\affiliation{Advanced Light Source, Lawrence Berkeley National Laboratory, Berkeley, CA 94720, USA}

\author{Donghui Lu}
\affiliation{Stanford Synchrotron Radiation Lightsource, SLAC National Accelerator Laboratory, Menlo Park, California 94025, USA}

\author{Makoto Hashimoto}
\affiliation{Stanford Synchrotron Radiation Lightsource, SLAC National Accelerator Laboratory, Menlo Park, California 94025, USA}

\author{Junichiro Kono}
\affiliation{Department of Electrical and Computer Engineering, Rice University, Houston, Texas 77005, USA}
\affiliation{Department of Physics and Astronomy and Rice Center for Quantum Materials, Rice University, Houston, TX, 77005 USA}
\affiliation{Department of Material Science and NanoEngineering, Rice University, Houston, Texas 77005, USA}

\author{Alex Frano}
\affiliation{Department of Physics, University of California San Diego,La Jolla, CA 92093,USA}

\author{Robert J. Birgeneau}
\affiliation{Department of Physics, University of California at Berkeley, Berkeley, California 94720, USA}
\affiliation{Materials Sciences Division, Lawrence Berkeley National Laboratory, Berkeley, California 94720, USA}
\affiliation{Department of Materials Science and Engineering, University of California, Berkeley, USA}

\author{Xiaodong Xu}
\affiliation{Department of Physics, University of Washington, Seattle, Washington 98195, USA}
\affiliation{Department of Materials Science and Engineering, University of Washington, Seattle,
Washington 98195, USA}

\author{Jian-Xin Zhu}
\affiliation{Theoretical Division and Center for Integrated Nanotechnologies, Los Alamos National Laboratory, Los Alamos, NM, USA}

\author{Pengcheng Dai}
\affiliation{Department of Physics and Astronomy and Rice Center for Quantum Materials, Rice University, Houston, TX, 77005 USA}

\author{Jiun-Haw Chu}
\affiliation{Department of Physics, University of Washington, Seattle, Washington 98195, USA}

\author{Ming Yi}
\email{mingyi@rice.edu}
\affiliation{Department of Physics and Astronomy and Rice Center for Quantum Materials, Rice University, Houston, TX, 77005 USA}

\date{\today}%

\begin{abstract}
Magnetism in two-dimensional (2D) materials has attracted considerable attention recently for both fundamental understanding of magnetism and its tunability towards device applications. The isostructural Fe$_3$GeTe$_2$ and \fgat~are two members of the Fe-based van der Waals (vdW) ferromagnet family, but exhibit very different Curie temperatures (\tc) of 210 K and 360 K, respectively. Here, by using angle-resolved photoemission spectroscopy and density functional theory, we systematically compare the electronic structures of the two compounds. Qualitative similarities in the Fermi surface can be found between the two compounds, with expanded hole pockets in Fe$_3$GaTe$_2$ suggesting additional hole carriers compared to Fe$_3$GeTe$_2$. Interestingly, we observe almost no band shift in \fgat~across its \tc~of 360 K, compared to a small shift in \fgt~across its \tc~of 210 K. The weak temperature-dependent evolution strongly deviates from the expectations of an itinerant Stoner mechanism. Our results suggest that itinerant electrons have minimal contributions to the enhancement of \tc~in \fgat~compared to \fgt, and that the nature of ferromagnetism in these Fe-based vdW ferromagnets must be understood with considerations of the electron correlations.

\end{abstract}

\maketitle


The recently discovered van der Waals (vdW) family of ferromagnets exhibits Curie temperatures (\tc) ranging from 30 K to above room temperature~\cite{Gong2017,Huang2017,Deng2018,Fei2018}. The remarkable preservation of long-range ferromagnetic order in these materials in the 2D regime position them as a promising class of materials for the development of next-generation spintronic devices~\cite{Yamada1998,Burch2018,Gong2019,Mak2019,Gibertini2019}. Equally importantly, these vdW materials offer a new platform to probe 2D magnetism. Our understanding of magnetism has been developed from two opposing limits. One approach is based on the weak-coupling picture where ferromagnetism arises from spontaneous spin splitting of the itinerant electronic bands near the Fermi level (\ef) onsetting at~\tc~\cite{Schmitt2012,Xu2020_ARPESnoStoner,Mattew2008,Stoner1938,Dordevic2001,MORIYA1979,Bloch1929,Slater1936,Edmund1947,Su2019}, which can only occur in metals. The other approach is based on a strong-coupling picture where electrons are localized and magnetism arises from the Heisenberg exchange coupling of the local moments, where the magnetic exchange splitting has no temperature dependence across \tc, and is often associated with insulators~\cite{Mattew2008,Schmitt2012,Williams2015,Chen2022_cgtneutron,Goodenough1967,Heisenberg1928}. In real materials, while the two limits exist, many compounds live in a regime where both mechanisms contribute. One such example is the iron-based superconductors (FeSCs), where electron correlations are moderate in between the strongly localized Mott physics of the cuprates and the itinerant spin-density-wave chromium metal. 
Neutron scattering identifies itinerant spin excitations at low energies with large fluctuating moments up to high energies~\cite{Dai2015}. Even contradictory reports of temperature-dependent exchange splitting have left a standing debate on the nature of magnetism in Fe and Ni metals~\cite{Eastman1978}.

\begin{figure*}
\includegraphics[width=\textwidth]{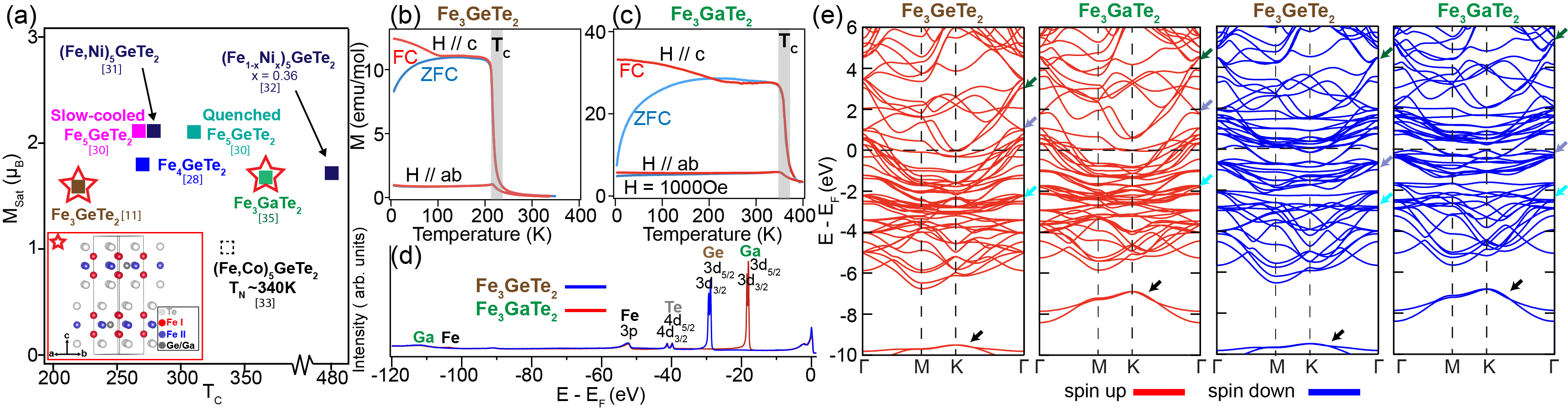}
\caption{
(a) \tc~and saturated moment (M$_{sat}$)~comparison for reported Fe-based vdW magnets. All are ferromagnets except (Fe,Co)$_5$GeTe$_2$, which is an antiferromagnet. The inset shows the crystal structure for Fe$_3$XTe$_2$ (X = Ge, Ga). (b-c) Temperature-dependent magnetization curves of Fe$_3$XTe$_2$. Zero-field-cooled (ZFC) and field-cooled (FC) curves indicate a ferromagnetic to paramagnetic transition at 210 K for Fe$_3$GeTe$_2$ and 360 K for Fe$_3$GaTe$_2$. (d) Core level photoemission spectra of Fe$_3$XTe$_2$. (e) DFT calculations for the ferromagnetic ground state, with the spin up and spin down bands indicated in red and blue, respectively. Spin-orbital coupling is not included. Pairs of arrows indicate the same features for comparison between the two compounds.
}\label{fig:Fig1}
\end{figure*}

The vdW ferromagnets can be largely grouped into two families, the insulating Cr-based compounds such as Cr$_2$Ge$_2$Te$_6$~\cite{Gong2017} and CrI$_3$ ~\cite{Huang2017}, and the metallic Fe-based compounds such as Fe$_n$XTe$_2$ (n = 3-5; X = Ge, Ga) (FGTs). The ferromagnetism in the insulating Cr-based compounds indeed can be understood by an anisotropic Heisenberg model where correlations between local moments persist to well above \tc~\cite{Williams2015,Chen2022_cgtneutron}. As a result, the electronic structure only exhibits very subtle evolution across \tc~\cite{Watson2020} while the FGTs are quite different. Consisting of Te-sandwiched vdW slabs, the various members of this family differ structurally in the number of Fe sites within each slab as well as the number of slabs within a unit cell dictated by the stacking order~\cite{Gong2019,Chen2013,Junho2020,Zhang2020,May2019,Stahl2018,Chen2022_1,Chen2022_2,Chen2023_F5,Zhang2022}. Notably, the \tc's of the FGTs are close to or even above room temperature  (Fig.~\ref{fig:Fig1}a)~\cite{Gong2019,Chen2013,Junho2020,Zhang2020,May2019,Stahl2018,Chen2022_1,Chen2022_2}. As metals, the FGTs are often referred to as itinerant magnets. However, ample evidence suggest a coexistence of both local moments and itinerant electrons. 
Fe$_{3-x}$GeTe$_2$ with a \tc~that varies between 140 K to 220 K~\cite{Deng2018,Fei2018}, in particular, has been demonstrated by neutron scattering to exhibit a dual nature of magnetic excitations~\cite{Bao2022}. Angle-resolved photoemission spectroscopy (ARPES) measurements indicate a deviation from Stone-type spin splitting across \tc~as well as spectral weight transfer suggestive of Kondo behavior~\cite{Xu2020_ARPESnoStoner,Zhang2018_Kondo,Zhu2016_heavyfermion,Bai2022,Kim2022}. Very recently, Fe$_3$GaTe$_2$, isostructural to Fe$_3$GeTe$_2$, has been synthesized and shown to exhibit a remarkable above-room temperature \tc~of 360 K, along with a high saturation magnetic moment, significant perpendicular magnetic anisotropy energy density, and a large anomalous Hall angle at room temperature~\cite{Zhang2022,Li2023,Jin_2023,jin2023roomtemperature,li2023roomtemperature}. These findings highlight the potential of Fe$_3$GaTe$_2$ as an exciting material for applications. The identical crystal structure yet drastically different \tc's in these two compounds offer an opportunity to probe into the nature of the magnetism in these materials.
Here, via systematic ARPES measurements and density functional theory calculations, we compare and contrast the electronic structure of \fgat~and \fgt. We find \fgat~to be an effectively hole-doped version of \fgt. In a large energy range of the valence bands, we identify a separation of the spectral weight that seems to be consistent with the predicted Fe spin up and spin down states. However, we find no observable shift in the electronic structures of \fgat~across its \tc, compared to a subtle shift for \fgt. Taken all together, the origin of magnetism in both Fe$_3$GaTe$_2$ and Fe$_3$GeTe$_2$ deviate strongly from the expectations of the itinerant Stoner model, with \fgat~exhibiting an even stronger local moment behavior. Our results indicate that the local moments are crucial for explaining the nature of ferromagnetism in FGTs, and are likely responsible for the much enhanced \tc~in \fgat.

\begin{figure*}
\includegraphics[width=\textwidth]{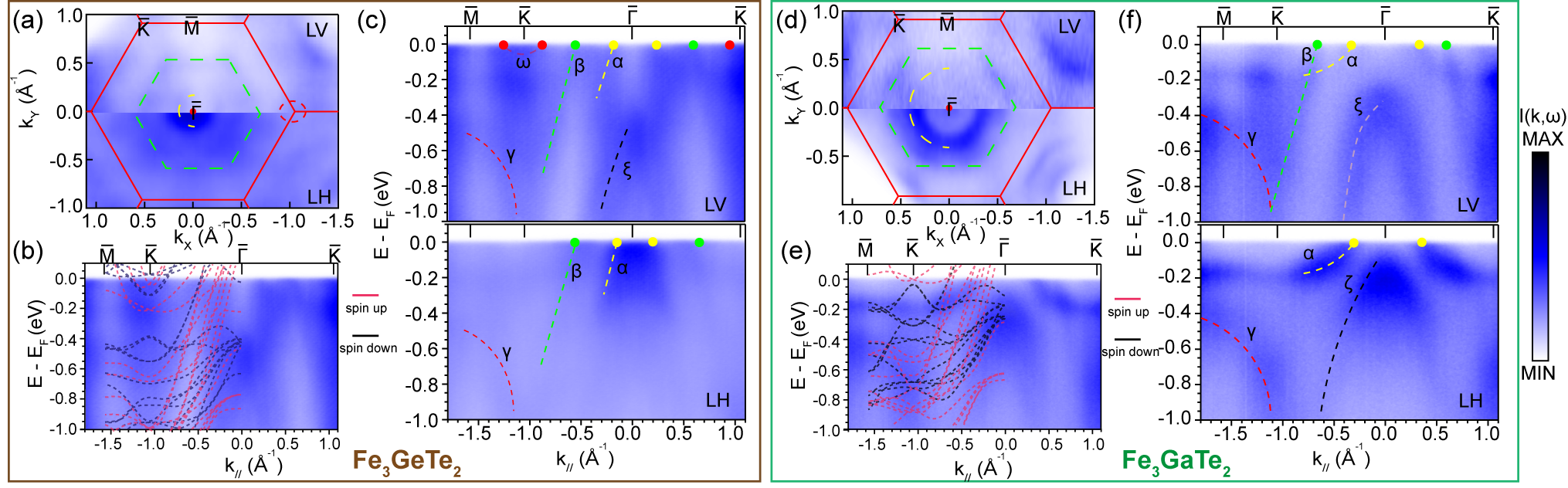}
\caption{
(a) Fermi surface mapping with LH and LV polarized light for Fe$_3$GeTe$_2$, with BZ boundaries labeled by red lines. (b) Measured dispersions along the $\overline{M}$-$\overline{K}$-$\overline\Gamma$-$\overline{K}$ direction overlaid with DFT calculations that are renormalized by a factor of 1.6. (c) LV and LH polarization dependent dipsersions measured along the $\overline{M}$-$\overline{K}$-$\overline\Gamma$-$\overline{K}$ direction, with key features marked by dashed lines. (d)-(f) Same measurements as (a)-(c) but for \fgat.
}\label{fig:Fig2}
\end{figure*}

High-quality \fgat~and \fgt~single crystals were synthesized by a chemical transport method~\cite{Bao2022}. ARPES measurements were carried out at beamline 5-2 of the Stanford Synchrotron Radiation Lightsource, ESM (21ID-I) beamline of the National Synchrotron Light Source II, and beamlines 7.0.2.1, 10.0.1 and 4.0.3 of the Advanced Light Source, using a DA30, DA30, R4000, and R8000 electron analyzer, respectively. The overall energy and angular resolutions were 15 meV and 0.1°, respectively. All data shown in the main text were taken with 132 eV photons, with additional photon energy-dependence data shown in the SM. All data were taken at 15 K unless otherwise noted. The DFT calculations were carried out by using WIEN2k package which uses the full-potential augmented plane wave plus local orbital as the basis~\cite{Blaha2020}. Perdew-Burke-Ernzefhof (PBE) generalized gradient approximation (GGA) was employed for the exchange-correlation functional~\cite{Perdew1996} and a 16 x 16 x 3 k-point mesh for self-consistent calculation.


As depicted in Fig. 1a, Fe$_3$XTe$_2$ has a layered hexagonal crystal structure in the space group P63/mmc (No. 194)~\cite{Kim2018}. The lattice parameter for Fe$_3$GaTe$_2$ (a = 4.07 \AA, c = 16.1 \AA) and Fe$_3$GeTe$_2$ (a = 3.99 \AA, c = 16.3 \AA) are similar, as previously reported~\cite{Kim2018,Zhang2022}. 
The unit cell consists of two vdW slabs, each with two nonequivalent Fe sites, Fe I and Fe II, and has the symmetry operators C$_{3z}$, C$_{2y}$ and P (inversion) that enforce the emergence of topological nodal lines in the presence of ferromagnetism, giving rise to a tunable intrinsic anomalous Hall current~\cite{Kim2018,Wu2022}. Our DFT calculations for the FM phase of both compounds are shown in Fig.~\ref{fig:Fig1}e, where the topological crossings at the K point can be seen. The zero-field-cooled (ZFC) and field-cooled (FC) magnetization measurements (Fig.~\ref{fig:Fig1}b,c) for the two compounds show a \tc~of 210 K for Fe$_3$GeTe$_2$ and 360 K for Fe$_3$GaTe$_2$, in agreement with previous reports~\cite{Deng2018,Fei2018,Xu2020_ARPESnoStoner,Zhang2022,Zhu2016_heavyfermion}. From a comparison of our DFT calculations for the ferromagnetic ground state, \fgat~is an effective hole-doped version of \fgt, as the band structure is qualitatively similar except a shifting down of the chemical potential in \fgat~(Fig.~\ref{fig:Fig1}e).

\begin{figure*}
\includegraphics[width=\textwidth]{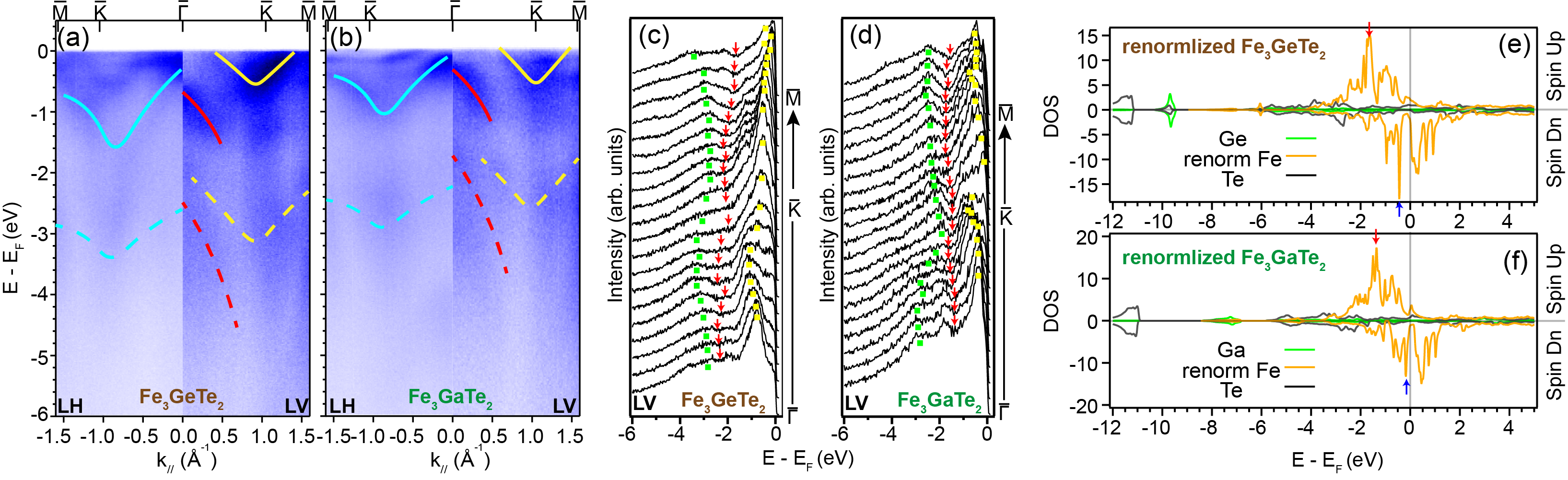}
\caption{(a)-(b) Large energy range spectra under both LH and LV polarizations for Fe$_3$GeTe$_2$ and \fgat. The solid lines highlight the coherent features near \ef~and the dashed lines highlight the corresponding incoherent features. (c)-(d) EDCs along the $\overline{M}$-$\overline{K}$-$\overline\Gamma$ directions for the LV polarization from (a)-(b). The green dots track the incoherent peaks and the yellow dots track the coherent peaks. The two features are separated by the dips that marked with red arrows. (e)-(f) Projected partial DOS calculated by DFT for Fe$_3$GeTe$_2$ and Fe$_3$GaTe$_2$. Only the Fe partial DOS is renormalized by a factor of 1.6. The arrows are pointing at the same features in \fgat~and \fgt, suggesting \fgat~to be a hole-doped version of \fgt.
}\label{fig:Fig3}
\end{figure*}

First, we show the integrated core-level photoemission spectrum of the two compounds in Figure~\ref{fig:Fig1}d. We also observe clear x-ray magnetic circular dichroism signal at the Fe L-edge for both compounds (see SM). As expected, the core level spectra for the two compounds are very similar except the distinct Ge and Ga 3d peaks. Next, we present the electronic structure of the ferromagnetic phase of Fe$_3$GaTe$_2$ and Fe$_3$GeTe$_2$ as measured by ARPES. The electronic structure of the two compounds near \ef~are compared in Fig.~\ref{fig:Fig2}, as measured by both linear vertical (LV) and linear horizontal (LH) polarized light. Consistent with previous reports on \fgt~\cite{Xu2020_ARPESnoStoner}, we observe two hole Fermi pockets centered at the $\overline\Gamma$ point: an inner circular pocket and an outer hexagonal pocket (Fig.~\ref{fig:Fig2}a). They are formed by two dispersive bands as observed on the high symmetry cut, and we label them as the $\alpha$ and $\beta$ bands (Fig.~\ref{fig:Fig2}c). 
Additionally, a small electron pocket is observed at the $\overline{K}$ point, which we label as the $\omega$ band. From the high symmetry cut, we also observe two other dispersions that do not cross \ef, which we label the $\xi$ and $\gamma$ bands.
For \fgat, we also observe two hole Fermi pockets at the $\overline\Gamma$ point (Fig.~\ref{fig:Fig2}d), both with similar shapes but expanded areas as compared to those in \fgt, indicating additional hole charge carriers in \fgat~compared to \fgt.
From the high symmetry cut (Fig.~\ref{fig:Fig2}f), both the inner $\alpha$ band and the outer $\beta$ band appear to cross \ef~at larger Fermi momenta. The $\gamma$ band is also observed to shift up in energy compared to that in \fgt. When we compare the near-\ef~measured dispersions with those by DFT calculations, we find that a renormalization factor of 1.6 can achieve a reasonable agreement for both compounds (Fig.~\ref{fig:Fig2}b,e), including the locations of the hole band tops at $\overline\Gamma$ (see SM for discussion). The renormalization factor is consistent with that determined for \fgt~previously~\cite{Xu2020_ARPESnoStoner}, and is slightly larger than that for Fe metal~\cite{Lichyenstein2001,SnchezBarriga2009Strength}.


Having examined the electronic structure in the near \ef~region, we next present the spectra in the large energy range covering the entire valence bands. Figure~\ref{fig:Fig3}a-b shows the spectra within 6 eV below \ef~along the $\overline{M}$-$\overline{K}$-$\overline\Gamma$ high symmetry direction for both compounds. Visibly, the spectrum is separated into sharp dispersions within 1 eV of \ef~and broad spectral intensity in the -2 to -3 eV energy range. As shown in Fig.~\ref{fig:Fig3}a-b, the sharp dispersions marked by the solid lines and the broad spectral intensity indicated by dashed lines exhibit similar dispersion patterns. This is also clearly shown in the stack of energy distribution curves (EDCs) in Fig.~\ref{fig:Fig3}c-d. The broad hump (green markers) largely follows the dispersion and photoemission matrix elements of the sharper bands near \ef~(yellow markers), and a clear dip (red arrows) separates the two regimes of sharp quasiparticles and broad spectral weight. Even in this large energy window, it is evident that the overall spectral shape of \fgat~is shifted up in energy compared to that of \fgt, consistent with the overall hole-doping. To understand the origin of these states, we look at the DFT calculated density of states (DOS) in the ferromagnetic state. Clearly, Fe 3$d$ states dominate the valence bands, with a small contribution by Te and Ge/Ga. To take into consideration the renormalization of the Fe 3$d$ states derived above, we renormalize the Fe partial DOS by 1.6 while leaving the Te and Ge/Ga partial DOS unrenormalized. This results in the spin majority (up) states having a peak near -2 eV and the spin minority (down) states near \ef. This comparison suggests that the sharper quasiparticles and broad hump dichotomy is likely dominated by the spin minority and spin majority states, respectively. Such kind of quasiparticle-dip-broad hump spectral feature has also been reported in other correlated ferromagnets, such as SrRuO$_3$, which is a metallic ferromagnet where both itinerant electrons and local moments contribute to the magnetism. In that case, this quasiparticle-dip-broad hump spectral feature was also explicitly reported, where strong scattering results in the incoherence of the spin majority states~\cite{Shai2013,Hahn2021}. In Fe and Ni metals, LDA+Dynamical Mean Field Theory has also captured such spectral lineshape in the single-particle spectral function by including the many-body effects of the 3$d$ states~\cite{Lichyenstein2001,Grechnev2007,Zhu2016_heavyfermion}.

\begin{figure*}
\includegraphics[width=\textwidth]{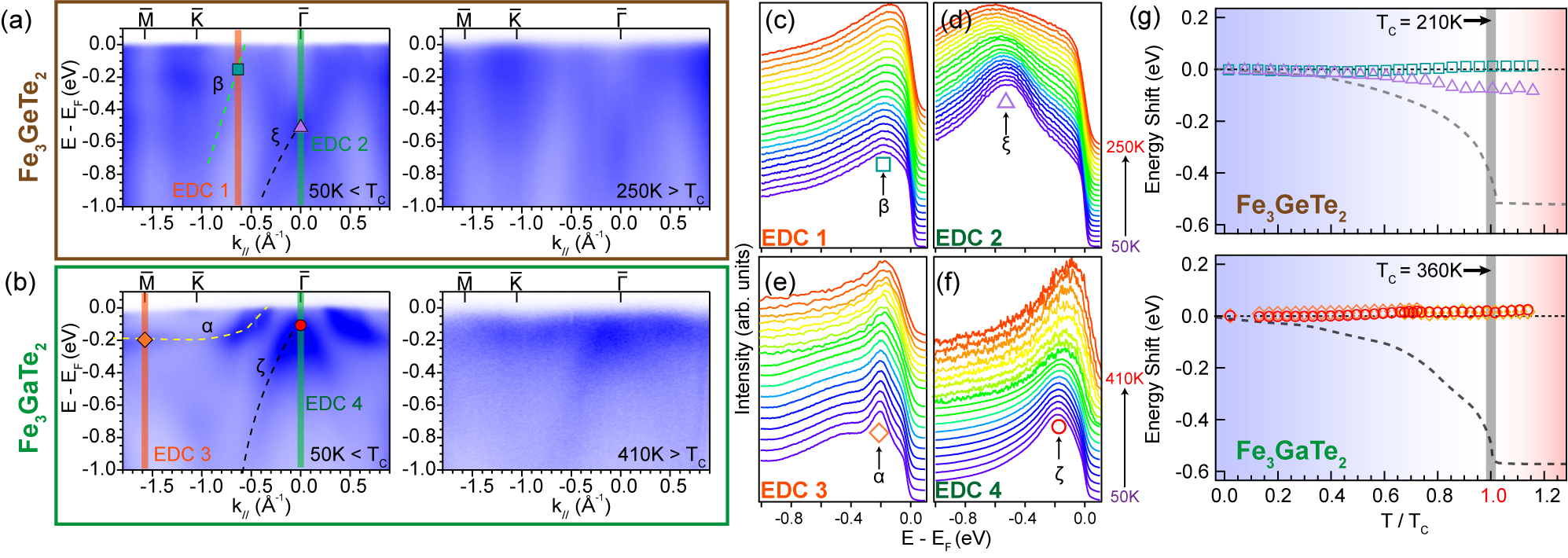}
\caption{
(a)-(b) Band dispersions in Fe$_3$GeTe$_2$ and Fe$_3$GaTe$_2$ above and below their respective \tc's. The main features are highlighted with dashed lines. (c)-(f) Temperature dependent EDCs, taken at the momenta marked in (a)-(b). (g) Fitted peak positions from the EDCs as a function of temperature, with the markers defined as in (c)-(f). The grey dashed lines indicate the estimated band shift from a Stoner model, determined by the DFT exchange splitting and magnetic moment for \fgt~\cite{Xu2020_ARPESnoStoner} and magnetization for \fgat~\cite{Zhang2022}.
}\label{fig:Fig4}
\end{figure*}

To examine the role of the near-\ef~bands in the ferromagnetism, we carried out temperature dependence study of the electronic structures for both Fe$_3$GaTe$_2$ and Fe$_3$GeTe$_2$ across their respective \tc. Figure~\ref{fig:Fig4}a shows the band dispersions of \fgt~(\tc~= 210 K) at 50 K and 250 K, with additional intermediate temperatures shown in the SM. And figure~\ref{fig:Fig4}b displays the band dispersion of Fe$_3$GaTe$_2$ (\tc~= 360 K) along the $\overline{M}$-$\overline{K}$-$\overline\Gamma$ direction at selected temperatures 50 K and 410 K, with additional intermediate temperatures shown in the SM. From those datasets, we can extract the temperature evolution of the bands by extracting the EDCs at specific momenta indicated by the orange and green lines in Fig.~\ref{fig:Fig4}(a) and Fig.~\ref{fig:Fig4}(b). The detailed EDCs are shown in Fig.~\ref{fig:Fig4}(c-f). For Fe$_3$GeTe$_2$, the $\beta$ band is observed to not shift, a small shift of the $\xi$ band is observed across \tc, consistent with previous report~\cite{Xu2020_ARPESnoStoner}. Similar  analysis was also carried out for \fgat. And for Fe$_3$GaTe$_2$, there is no noticeable shift in the peak position of the EDCs for both $\alpha$ and $\zeta$ bands with increasing temperature through the \tc~of 360 K, as shown also from the fitted peak positions in Fig.~\ref{fig:Fig4}(g). The lack of observable shift of bands across \tc~strongly deviates from the expected behavior of itinerant ferromagnets. According to the Stoner model, the exchange splitting is expected to disappear above \tc~\cite{Huang2021}. For Fe$_3$GeTe$_2$ and Fe$_3$GaTe$_2$, we can estimate the exchange splitting sizes from DFT calculations to approximately be 1.5 eV and 1.7 eV, respectively. The expected shift of majority/minority bands would be half of the exchange splitting divided by the band renormalization factors, resulting in 0.47 eV and 0.53 eV, respectively. The temperature evolution of the band shift according to the Stoner model can then be estimated by scaling this energy scale to the existing temperature-dependent bulk magnetization in \fgat~\cite{Zhang2022} or magnetic moment measured by neutron diffraction for \fgt~\cite{Xu2020_ARPESnoStoner,Zhang2022}, which are plotted as the grey dashed lines in Fig.~\ref{fig:Fig4}g. We note that for systems whose ferromagnetism is contributed by both itinerant electrons and local moments, partial closing of the exchange splitting is observed across \tc, such as Fe metal~\cite{Hubbard1979,Isabelle1994}, MnSi~\cite{Fang2022,Jin2023} and SrRuO$_3$~\cite{Shai2013,Hahn2021}. The stark contrast between the expected shift and the observed band shift here suggests that itinerant electrons play a minimal role in the ferromagnetism in Fe$_3$GeTe$_2$ and Fe$_3$GaTe$_2$. While some finite shift is still observed in \fgt, we observe no shift in \fgat, suggesting that local moments play an even more dominant role in \fgat. 

Taking all the presented evidence together, we come to an understanding of the ferromagnetism of the isostructural Fe$_3$GaTe$_2$ (\tc $\sim$ 360 K) and Fe$_3$GeTe$_2$ (\tc $\sim$ 210 K) as the following: while both systems are metallic and exhibit clear Fermi surfaces, the valence band spectral intensity exhibit a quasiparticle-dip-broad hump feature that seem to indicate non-negligible correlation effects. In addition, the itinerant charge carriers near \ef~show minimal modifications across \tc, with \fgat~showing even less observable changes. This indicates that the large enhancement of \tc~in \fgat~can not barely be due to the change in the itinerant charge carriers and must be a result of the local moments. Our findings therefore demonstrates that the \fgat~and \fgt~systems are moderately correlated and a comprehensive understanding of the magnetism in these Fe-based vdW ferromagnets must take into consideration the many-body interactions of the Fe 3$d$ states.

This research used resources of the Advanced Light Source, and the Stanford Synchrotron Radiation Lightsource, both U.S. Department Of Energy (DOE) Office of Science User Facilities under contract nos. DE-AC02-05CH11231 and AC02-76SF00515, respectively. ARPES work is supported by the U.S. DOE grant No. DE-SC0021421, the Gordon and Betty Moore Foundation’s EPiQS Initiative through grant no. GBMF9470. YCG is supported by the Robert A. Welch Foundation, Grant No. C-2175 (M.Y.).
Work at University of California, Berkeley, is funded by the U.S. Department of Energy, Office of Science, Office of Basic Energy Sciences, Materials Sciences and Engineering Division under Contract No. DE-AC02-05-CH11231 (Quantum Materials program KC2202). 
Work at Los Alamos was carried out under the auspices of the U.S. Department of Energy (DOE) National Nuclear Security Administration (NNSA) under Contract No. 89233218CNA000001, and was supported by LANL LDRD Program and in part by Center for Integrated Nanotechnologies, a DOE BES user facility, in partnership with the LANL Institutional Computing Program for computational resources. Materials synthesis at UW was supported as part of Programmable Quantum Materials, an Energy Frontier Research Center funded by the U.S. Department of Energy (DOE), Office of Science, Basic Energy Sciences (BES), under award DE-SC0019443.

\bibliography{bib2}

\end{document}